\documentclass[journal]{IEEEtran}

\usepackage{cite}
\usepackage{bm}
\usepackage{amsmath}
\usepackage{graphicx}
\usepackage{amssymb}
\usepackage{multirow}
\usepackage{array}
\usepackage{color}
\usepackage{epsfig}
\usepackage{graphics} 
\usepackage{balance,epsfig} 
\usepackage{algorithmic}

\allowdisplaybreaks

\renewcommand{\algorithmicensure}{Output:}
\renewcommand{\algorithmicrequire}{Input:}

\newtheorem{theorem}{Theorem}

\newtheorem{example}{Example}

\newcommand{\cN}{\mathcal{N}}
\newcommand{\cI}{\mathcal{I}}
\newcommand{\cO}{\mathcal{O}}
\newcommand{\cF}{\mathcal{G}}
\newcommand{\cR}{\mathcal{R}}

\newcommand{\lfteqn}{\begin{eqnarray} \begin{array}{lllllll}}
		\newcommand{\ndeqn}{\end{array} \nonumber \end{eqnarray}}
\newcommand{\Lfteqn}{\begin{eqnarray} \begin{array}{lllllll}}
		\newcommand{\Ndeqn}{\end{array}  \end{eqnarray}}
\def\bull{\vrule height 1.1ex width 1.1ex depth -.0ex }

\begin{document}
	
\title{Learning in Dynamic Systems and Its Application to Adaptive PID Control}

\author{Omar Makke and Feng Lin  
	\thanks{This work is supported in part by the National Science Foundation of USA under grant 2146615.}
	\thanks{Omar Makke is technical expert at Ford Motor Company. Feng Lin is with the Department of Electrical and Computer Engineering, Wayne State University, Detroit, MI 48202, USA. E-mail: omarmakke@wayne.edu and flin@wayne.edu.} }

\maketitle \thispagestyle{empty} \pagestyle{empty}

\begin{abstract}
	
Deep learning using neural networks has revolutionized machine
learning and put artificial intelligence into everyday life. In
order to introduce self-learning to dynamic systems other than
neural networks, we extend the Brandt-Lin learning algorithm of
neural networks to a large class of dynamic systems. This
extension is possible because the Brandt-Lin algorithm does not
require a dedicated step to back-propagate the errors in neural networks. To this
end, we first generalize signal-flow graphs so that they can be
used to model nonlinear systems as well as linear systems. We then
derive the extended Brandt-Lin algorithm that can be used to adapt
the weights of branches in generalized signal-flow graphs. We show
the applications of the new algorithm by applying it to adaptive
PID control. In particular, we derive a new adaptation law for PID
controllers. We verify the effectiveness of the method using
simulations for linear and nonlinear plants, stable as well as unstable plants.

	
\end{abstract}

\section{Introduction}

Machine learning and artificial intelligence are being adopted in
increasingly more engineering fields. They have made significant
impacts in many engineering fields by solving problems that were considered difficult in the past. However, the impact of
machine learning and artificial intelligence on control systems
are limited in comparison to their impact in other fields. One
reason for this limited impact is that the learning algorithms
used in machine learning and artificial intelligence are not
directly applicable to control systems. This is unfortunate
because control is one of the first fields where
adaptation/learning were proposed. Adaptive control has been
developed and used for many decades. Significant results have been
obtained in adaptive control. These results have been applied to
solve some difficult control problems.

In order to bridge the gap between adaptive control and new
machine learning techniques, we propose a new learning algorithm
that is inspired by the back-propagation learning algorithm in
neural networks. This new algorithm can be used in a wide range of
dynamic systems, including linear and nonlinear systems. We apply
this new algorithm to adaptive PID control and develop a new
method for adapting a PID controller.

To put our approach in proper context, let us first review some
results in machine learning and artificial intelligence. Machine
learning as a part of artificial intelligence has been
investigated as early as the 1950s \cite{samuel1959some}
\cite{nilsson1965learning}. Since then, many results on machine
learning have been developed \cite{michie1994machine}
\cite{jordan2015machine} \cite{mohri2018foundations}
\cite{sammut2011encyclopedia}. Among various methods of machine
learning, deep learning using neural networks is one of the most
widely use method \cite{widrow1994neural}
\cite{jain1996artificial} \cite{mandic2001recurrent}
\cite{gurney2014introduction}. In the past ten years or so, deep
learning using various neural networks has revolutionized the
fields of speech recognition, object recognition and detection
\cite{goodfellow2016deep} \cite{deng2014deep}
\cite{schmidhuber2015deep}. This achievement is partly due to the
availability of fast computers and large collections of data.

Although various learning algorithms\footnote{The word
	``algorithm'' is used here in a generalized sense to mean a model
	or a mathematical description for updating weights/parameter of
	neural networks or dynamic systems.} have been proposed for neural
networks, the back-propagation algorithm is probably the most well
known and widely used \cite{rumelhart1986learning}
\cite{hecht1992theory} \cite{chauvin2013backpropagation}. It
allows errors to be back propagated via a feedback network so that
the strengths of synapses (or weights of connections) can be
adapted to reduce a given performance index.

In \cite{brandt1996supervised} \cite{lin2000self}
\cite{brandt1999adaptive} \cite{lin2020supervised}, Brandt and Lin
developed a learning algorithm that is mathematically equivalent to
the back-propagation algorithm in neural networks but have the
following advantages over the back-propagation algorithm. (1) It
does not require a dedicated feedback step to back-propagate errors. This
makes its implementations, especially implementations on silicon,
much simpler. (2) It is biologically plausible as all information
needed for synapses to adapt is available in a biological neuron.
Hence, artificial neural networks can indeed mimic biological neural networks. (3) As to be shown in this paper, it
can be extended for use in general dynamic systems, that is, many
dynamic systems can adapt in a way similar to that of neural
networks. This property allows us to introduce learning capability
into a large class of engineering systems.

In this paper we extend the Brandt-Lin algorithm to dynamic systems. We first
generalize conventional signal-flow graphs (CSFG) to generalized
signal-flow graphs (GSFG). A GSFG has all the elements of CSFG, and in
addition, some nodes in GSFG are super nodes that can have (linear
or nonlinear) dynamics, described by transfer functions or (linear
and nonlinear) differential equations. In this formulation, each super
node represents a subsystem that is connected via branches.
Signals flow between nodes the same way as in CSFG, that
is, the input signal to a node is the sum of signals from all
branches leading to the node. Gains of some branches are adaptive,
that is, they can be adapted in a way similar to the adaptation of
strengths of synapses in a neural network. GSFG can be used to
model a large class of dynamic systems, because it allows
nonlinear dynamics. To the best of our knowledge, no one has proposed to use GSFG to model a dynamic system and no one has developed an algorithm to learn general dynamic systems.

We extend the Brandt-Lin learning algorithm to dynamic systems
that can be modeled by GSFG as follows. We first partition the set
of the branches into adaptive branches and non-adaptive branches. The
goal of the learning is to minimize a cost function (or error) $E$ by
adapting the gains of adaptive branches. Denote the gain of the
branch from node $i$ to node $j$ by $w_{ij}$. We calculate the
derivatives of $E$ with respect to $w_{ij}$ and use gradient
decent to adapt $w_{ij}$ of adaptive branches as $\dot{w}_{ij}=
-\gamma dE/dw_{ij}$ for some adaptation parameter $\gamma >0$. We
show that the $\dot{w}_{ij}$ of a branch entering a node $j$ can
be expressed in terms of $\dot{w}_{jm}$ of all branches leaving
the node $j$, where $m$ is a node connected from node $j$. We further show that this relationship is linear and
derive a set of linear equations describing this relationship,
which gives us the extended Brandt-Lin algorithm. As long as the
set of linear equations has a solution, which is guaranteed by the
corresponding determinant being non-zero, $E$ can be reduced using
the extended Brandt-Lin algorithm.

Although the extended Brandt-Lin algorithm can be used in many
dynamic systems, we focus on control systems in this paper. We
propose to use the extended Brandt-Lin algorithm for model
reference adaptive PID control. The goal of model reference
adaptive control is to adapt parameters in a controller such as
the PID gains so that the closed-loop system approaches a given
reference model. We model the closed-loop system using GSFG and
represent the adaptive parameters as gains of some adaptive
branches in the GSFG. The extended Brandt-Lin algorithm can then
be used to adapt these gains to achieve model reference adaptive
control. In this approach, the closed loop system does not have to match the reference model in structure.

We apply this method to adaptive PID control and derive the
adaptation law for the PID gains using the extended Brandt-Lin
algorithm. We implement the adaptive PID controller in Simulink
and test the effectiveness of the method for various types of
plants (systems to be controlled), including stable and unstable
plants, linear and nonlinear plants. Simulation results show that
the method works very well.

Adaptive control has been used extensively in control systems
\cite{aastrom2013adaptive} \cite{landau2011adaptive}
\cite{tao2003adaptive}. It has been applied to practical problems
such as flight control \cite{singh1996adaptive}, mobile robots
\cite{wang2014distributed}, motion control \cite{sun2016transient}
and others \cite{aastrom1983theory}. Model reference adaptive
control is also investigated extensively in the literature
\cite{goodwin1987parameter} \cite{kreisselmeier1982stable}
\cite{yu1996model}. The applications of model reference adaptive
control include robotic manipulators \cite{zhang2017review}, wind
energy systems \cite{mosaad2018model}, motor drive systems
\cite{nguyen2018model}, and unmanned underwater vehicle
\cite{makavita2016predictor}. Obviously, the results in this paper
are different than those in the literature.

The paper is organized as follows. In Section II, we give a brief
review of the Brandt-Lin algorithm for general neural networks,
including both hierarchical networks and non-hierarchical
networks. In Section III, we introduce generalized signal-flow
graphs by introducing super nodes in conventional signal-flow
graphs. We use functionals to describe
dynamics of super nodes. In Section IV, we derive the extended
Brandt-Lin algorithm, which can be implemented on-line. For
dynamic systems with feedbacks, we provide a necessary and
sufficient condition for the existence of a unique solution to the
algorithm. In Section V, we show how to calculate Fr\'{e}chet
derivatives, which are needed in the extended Brandt-Lin
algorithm. We consider both linear dynamics and nonlinear
dynamics. In Section VI, we apply the extended Brandt-Lin
algorithm to model reference adaptive control by considering an
adaptive PID controller. We derive the adaptation law for the PID
controller. We then evaluate the effectiveness of the PID
controller using Simulink for stable and unstable plants, as well
as linear and nonlinear plants.

\section{Brandt-Lin Learning Algorithm}

Because the new learning algorithm to be proposed is an extension
of the Brandt-Lin learning algorithm from neural networks to
general systems, let us briefly review the Brandt-Lin algorithm.

To describe a neural network (either hierarchical or
non-hierarchical), we enumerate all neurons in a neural network as
$\cN = \{1, 2, ..., N \}$. We do not put any restrictions on
connections among neurons. The weights of the connection from the
$i$-th neuron  to the $j$-th neuron is denoted by $w_{ij}$. The
set of all connections is denoted by
$$
\Psi = \{w _{ij}: i, j \in \cN \wedge i \mbox{ is connected to } j
\}.
$$

Not all neurons have preceding neurons. If a neuron does not have
preceding neutrons, then we consider it as an input neuron. The
set of input neurons is denoted by
$$
\cI = \{n \in \cN: (\forall j \in \cN) w_{jn} \not\in \Psi \}.
$$
The firing rates of input neurons $r_n$, $n \in \cI$, are considered as the inputs to the neural network.

The dynamics of non-input neuron $n \in \cN - \cI$ are described
by its membrane potential $r_n$ and firing rate $r_n$. The
membrane potential of the $n$-th neuron is the weighted sum of the
firing rates of its preceding neurons:
\begin{equation} \label{p}
	\begin{split}
		p_n = \sum _{w _{mn} \in \Psi} w _{mn} \ r_m .
	\end{split}
\end{equation}
The firing rate of the $n$-th neuron is given by
\begin{equation} \label{r}
	\begin{split}
		r_n = \sigma (p_n) ,
	\end{split}
\end{equation}
where $\sigma (p_n) = 1/(1+e^{-p_n})$ is the sigmoidal function.

The weights $w_{ij}$ can be adapted to minimize the following
least square error
$$
E=\frac{1}{2} \sum _{m \in \cO} (r_m - \bar{r}_m)^2,
$$
where $\cO$ is the set of output neurons and $\bar{r}_m$ is the
desired/target firing rate of the output neuron $m \in \cO$.

The following learning algorithm is proposed by Brandt and Lin in
\cite{brandt1996supervised} \cite{lin2020supervised} to adapt the
weights $w_{ij} \in \Psi$.
\begin{equation} \label{BLAlg}
	\begin{split}
		\dot{w}_{ij} & = \sigma ' (p_j) \frac{r_i }{r_j} (-\gamma r_j
		(r_j-\bar{r}_j) + \sum _{w _{jm} \in \Psi} w _{jm} \
		\dot{w}_{jm}),
	\end{split}
\end{equation}
where $\sigma '(p_j)$ is the derivative of $\sigma (p_j)$.

If the above equation has a unique solution for $w_{ij} \in \Psi$,
then it is proved in \cite{brandt1996supervised}
\cite{lin2020supervised} that the following is true.
\begin{equation} \label{GDAdap}
	\begin{split}
		\dot{w}_{ij} & = -\gamma \frac{dE}{dw_{ij}} ,
	\end{split}
\end{equation}
that is, the gradient-decent-based learning is achieved. Note that the significance of equation \ref{BLAlg} is that the adaptation of the weights is described as a function of time, which makes it suitable for online learning.

It is shown in \cite{brandt1996supervised}
\cite{lin2020supervised} that the above Brandt-Lin learning
algorithm has the following properties.

\begin{enumerate}
	\item
	The Brandt-Lin algorithm is mathematically equivalent to the
	well-known back-propagation algorithm. In other words, the
	Brandt-Lin algorithm can be used wherever the back-propagation
	algorithm can be used.
	\item
	The implementation of the Brandt-Lin algorithm does not require a
	dedicated feedback step, and thus a feedback network. A feedback-network-free implementation is given
	in \cite{lin2020supervised}.
	\item
	It is more plausible that the adaptation according to the
	Brandt-Lin algorithm can occur in biological neural systems,
	because it does not require a feedback network. It is unlikely
	that a biological neural system will have a feedback network with
	the same topology and synaptic weights as the feed-forward
	network.
	\item
	Using the Brandt-Lin algorithm, the information needed for
	dendritic synapses to adapt is available in the weights of axonic
	synapses and their rates of change, that is, $\sum _{w _{jm} \in
		\Psi} w _{jm} \ \dot{w}_{jm}$ in Equation (\ref{BLAlg}).
	\item
	The removal of a feedback network also eliminates the needs for
	two-phase adaptation (a feed-forward phase to generate the outputs
	for given input stimulus and a feedback phase to adapt the synapse
	strengths according to the error feedback). Hence, adaptation can
	be performed in a phaseless fashion by processing information
	asynchronously and concurrently.
	\item
	The adaptation parameter $\gamma$ appears only at the output
	neurons and hence can be easily adjusted during the adaptation.
	\item
	For layered neural networks, all layers have same or similar
	structures. If all layers have same number of neurons, then all
	layers are identical, except the last layer. This makes
	implementation of neural networks much easier.
	\item
	The Brandt-Lin algorithm is much easier to implement on silicon,
	because there is no feedback network and hence no wiring between
	the feedback network and feed-forward network.
	\item
	The Brandt-Lin algorithm provides the potential for designing
	neural networks with dynamically reconfigurable topologies,
	because adaptive neurons can be implemented using identical and
	standard units that can be connected arbitrarily.
	\item
	The implementation of the Brandt-Lin algorithm are more
	fault-tolerant because failures of some neurons do not cause the
	entire neural network to become nonfunctional as all connections
	are local.
	\item
	As to be shown in this paper, the Brandt-Lin algorithm can be extended so that it can be used in general dynamic systems other than neural networks. This extension allows a large class of dynamic systems to adapt in a way similar to neural networks.
\end{enumerate}

In the rest of the paper, we investigate the extension to general
dynamic systems. We first propose generalized signal-flow graphs
to model general dynamic systems. We then develop the
generalization of Brandt-Lin algorithm for generalized signal-flow
graphs.

\section{Generalized Signal-Flow Graphs}

A generalized signal-flow graph has all elements of a conventional signal-flow graph. In addition, some nodes in GSFG are super nodes as to be discussed below.

Assume that a GSFG has $N$ nodes. Denote a node by
$$
n \in \cN = \{1, 2, ..., N \}.
$$
Denote the branch (if exists) and its gain from node $i$ to node $j$ by $\omega _{ij}$. The set of branches/gains is denoted by
$$
\Omega = \{\omega _{ij}: i, j \in \cN \wedge \mbox{ node } i
\mbox{ is connected to node } j \}.
$$
The set $\Omega$ is partitioned into two sets:
$$
\Omega = \Omega_a \cup \Omega_{na},
$$
where $\Omega_a$ is the set of adaptable branches/gains and
$\Omega_{na}$ is the set of non-adaptable branches/gains.
Non-adaptable branches have gains which are constants,  that is,
$$
\omega _{ij} \in \Omega_{na} \Leftrightarrow \omega _{ij} =
\bar{\omega} _{ij},
$$
where $\bar{\omega} _{ij}$ are constants.

As mentioned above, some nodes in $\cN$ are super nodes. A super
node consists of a pair of input and output, denoted by
$$
(u_n, y_n),
$$

\noindent
where $u_n$ is the input to node $n$ and $y_n$ is the output from
node $n$. Let $U = \{u_n: \cR \rightarrow \cR\}$ be a set of all inputs to a super node $n$.  The relationship between $u_n$ and $y_n$ is described by
$$
y_n(t) = \cF _n [ u_n(t)],
$$
where $\cF _n$\footnote{$\cF _n
	$ can be viewed as the model of a single-input-single-output
	system starting at $-\infty$.} is a functional which maps every input function of time to an output function of time. If the super node is linear time invariant, then $\cF _n$ is the convolution of the input with the impulse response of the super node.

We assume that the Fr\'{e}chet derivative of $\cF _n$
exists\footnote{The Fr\'{e}chet derivative
	\cite{luenberger1997optimization} of $\cF _n$ is denoted by $\cF'
	_n$ and defined as a functional such that
	$$
	\lim _{||\varepsilon|| \rightarrow 0} \frac{||\cF
		_n[u+\varepsilon] -\cF _n[u]-\cF' _n[u] \varepsilon||}
	{||\varepsilon||} = 0 .
	$$}.

If a node $n \in \cN$ is not a super node, then $y_n=u_n$, that
is, $\cF _n$ is an identity mapping: $\cF _n [ u_n(t)]=u_n(t)$.

As in CSFG, the input signal of node $n$ is the sum of all signals
flowing to $n$:
\begin{equation} \label{u}
	\begin{split}
		u_n = \sum _{m=1}^N \omega _{mn} \ y_m .
	\end{split}
\end{equation}

The output signal of node $n$ is then given by
\begin{align*}
	y_n = \left\{ \begin{array}{ll} \cF_n [ u_n ] & \mbox{if $n$ is a super node } \\
	u_n & \mbox{otherwise}
	\end{array} \right.
\end{align*}
If we let $\cF_n [ u_n ] = u_n$ be the identity mapping if $n$ is not a super node, then the above can be written uniformly as
\begin{equation} \label{y}
	\begin{split}
		y_n = \cF_n [ u_n ].
	\end{split}
\end{equation}

\begin{example}
	
	Let us illustrate GSFG by a control system using PID controller as
	shown in Figure 1. The system consists of 8 nodes, including 4
	super nodes (SN).
	
	SN 1 is the integral part with transfer function $1/s$.
	
	SN 2 is the proportional part with transfer function $1$.
	
	SN 3 is the derivative part with transfer function $s$.
	
	SN 4 is the system to be controlled (plant). If the plant is
	linear, then it is represented by a transfer function $G_4(s)$. If
	the plant is nonlinear, then it is represented by a functional
	$\cF _4$.
	
	The PID gains are represented by branch gains as follows.
	\lfteqn
	\mbox{Integral:} & K_I = \omega _{14} \\
	\mbox{Proportional:} & K_P = \omega _{24} \\
	\mbox{Derivative:} & K_D = \omega _{34}
	\ndeqn
	All other branch gains are 1, except $\omega _{46}=-1$.
	
	\begin{figure}[htb] \label{Figure1}
		\centering
		\includegraphics[keepaspectratio=true,angle=0,width=0.5\textwidth]{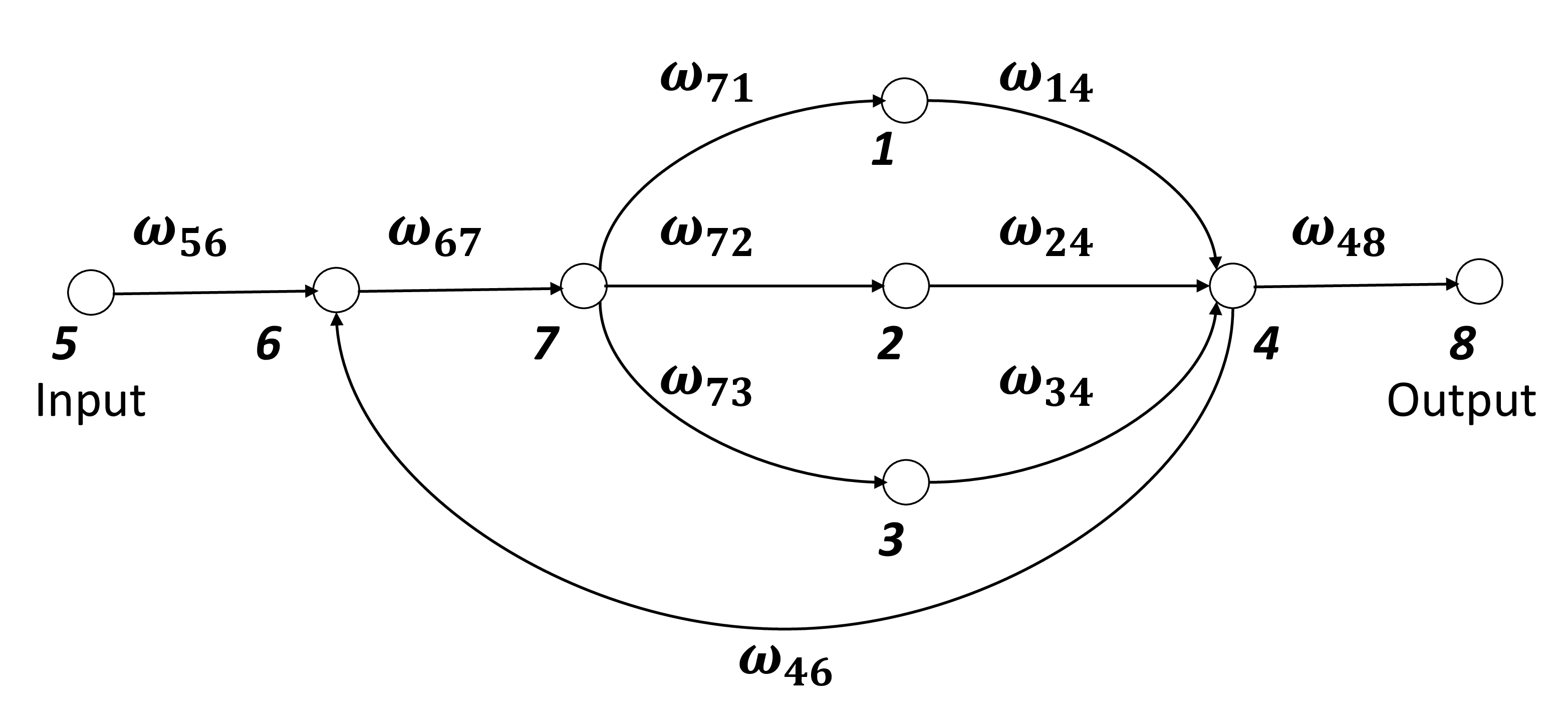}
		\caption{generalized signal-flow graph of a control system using
			PID controller}
	\end{figure}

	$\omega _{14}, \omega _{24}, \omega _{34}$ are adaptable branch gains. All other branch gains are non-adaptable.
	
\end{example}

\section{On-line Learning}

Our goal is to use on-line learning to learn/adapt the gains
$\omega _{ij} \in \Omega_a$ so that some error is minimized. We
assume that the error is a function of outputs:
$$
E=E(y_1, y_2,..., y_N).
$$
Not all $y_n$ appear explicitly in $E$. We can view nodes whose
outputs appear explicitly in $E$ as {\em output nodes} and denoted
the set of output nodes as $\cO$. One error often used is the
least square error
$$
E=\frac{1}{2} \sum _{m \in \cO} (y_m - \tilde{y}_m)^2,
$$
where $\tilde{y}_m$ is the desired/target output of node $m \in
\cO$.

To achieve on-line learning, we use gradient decent to learn the
gains ${\omega} _{ij} \in \Omega _a$ as
\begin{align*}
	\dot{\omega}_{ij} & = -\gamma \frac{dE}{d\omega_{ij}} .
\end{align*}
where $\gamma$ is the adaptation/learning rate, which is a design
parameter.

\begin{theorem} \label{Thm1}
	
	Consider an adaptive system described by a generalized
	signal-flow graph with nodes $n \in \cN$ and branches $\omega
	_{ij} \in \Omega$. Assume that the following set of equations for
	${\omega} _{ij} \in \Omega $ have a unique solution.
	\begin{equation} \label{Alg}
		\begin{split}
			\dot{\omega}_{ij} & = \cF'_j [u_j] \frac{y_i }{y_j} (-\gamma y_j
			\frac{\partial E}{\partial y_j} \\
			& + \sum _{\omega _{jm} \in \Omega_a} \omega _{jm} \
			\dot{\omega}_{jm}+ \sum _{\omega _{jm} \in \Omega_{na}}
			\bar{\omega} _{jm} \ \dot{\omega}_{jm}).
		\end{split}
	\end{equation}
	Then, using the above equation, the gradient-decent-based on-line
	learning is achieved as
	\begin{equation} \label{Adap}
		\begin{split}
			\dot{\omega}_{ij} & = -\gamma \frac{dE}{d\omega_{ij}} ,
		\end{split}
	\end{equation}
	
\end{theorem}
{\em Proof}

By the assumption, Equation (\ref{Alg}) has a unique solution.
Hence, we only need to show that Equation (\ref{Adap}) is a
solution to Equation (\ref{Alg}).

Clearly,
\begin{equation} \label{wjm}
	\begin{split}
		\frac{dE}{d \omega _{ij}} & = \frac{dE}{dy_j} \frac{dy_j}{du_j}
		\frac{du_j}{d \omega _{ij}} \\
		& = \frac{dE}{dy_j} \cF ' (u_j) \frac{du_j}{d \omega _{ij}} \\
		& (\mbox{by Equation (\ref{y}) with }n=j)\\
		& = \frac{dE}{dy_j} \cF ' (u_j) y_i \\
		& (\mbox{by Equation (\ref{u}) with $n=j$ and } m=i) .
	\end{split}
\end{equation}
On the other hand,
\begin{align*}
	\frac{dE}{dy_j} & = \frac{\partial E}{\partial y_j} + \sum
	_{\omega _{jm} \in \Psi} \frac{dE}{dy_m}
	\frac{dy_m}{du_m} \frac{du_m}{dy_j} \\
	& = \frac{\partial E}{\partial y_j} + \sum _{\omega _{jm} \in
		\Psi} \frac{dE}{dy_m} \cF '(u_m) \frac{du_m}{dy_j} \\
	& (\mbox{by Equation (\ref{y}) with }n=m)\\
	& = \frac{\partial E}{\partial y_j} + \sum _{\omega _{jm} \in
		\Psi} \frac{dE}{dy_m} \cF '(u_m) \omega_{jm} \\
	& (\mbox{by Equation (\ref{u}) with $n=m$ and } m=j) \\
	& = \frac{\partial E}{\partial y_j} + \sum _{\omega _{jm} \in
		\Psi} \frac{dE}{d \omega _{jm}} \frac{\omega_{jm}}{y_j} \\
	& (\mbox{by Equation (\ref{wjm}) with $i=j$ and }j=m)
\end{align*}

Hence, by Equation (\ref{wjm}),
\begin{align*}
	\frac{dE}{d \omega _{ij}} & = \frac{dE}{dy_j} \cF ' (u_j) y_i \\
	& = (\frac{\partial E}{\partial y_j} + \sum _{\omega _{jm} \in
		\Psi} \frac{dE}{d \omega _{jm}} \frac{\omega_{jm}}{y_j}) \cF '
	(u_j) y_i .
\end{align*}
By Equation (\ref{Adap}),
\begin{align*}
	\frac{dE}{d \omega _{ij}} & = \frac{\dot{\omega}_{ij}}{-\gamma} \\
	\frac{dE}{d \omega _{jm}} & = \frac{\dot{\omega}_{jm}}{-\gamma} .
\end{align*}

Therefore,
\begin{align*}
	& \frac{dE}{d \omega _{ij}} = (\frac{\partial E}{\partial y_j} +
	\sum _{\omega _{jm} \in \Psi} \frac{dE}{d \omega _{jm}}
	\frac{\omega_{jm}}{y_j}) \cF ' (u_j) y_i \\
	\Leftrightarrow & \frac{\dot{\omega}_{ij}}{-\gamma} =
	(\frac{\partial E}{\partial y_j} + \sum _{\omega _{jm} \in \Psi}
	\frac{\dot{\omega}_{jm}}{-\gamma} \frac{\omega_{jm}}{y_j}) \cF '
	(u_j) y_i \\
	\Leftrightarrow & \dot{\omega}_{ij} = (-\gamma \frac{\partial
		E}{\partial y_j} + \sum _{\omega _{jm} \in \Psi} \dot{\omega}_{jm}
	\frac{\omega_{jm}}{y_j}) \cF ' (u_j) y_i \\
	\Leftrightarrow & \dot{\omega}_{ij} = (-\gamma y_j \frac{\partial
		E}{\partial y_j} + \sum _{\omega _{jm} \in \Psi} \dot{\omega}_{jm}
	\omega_{jm}) \cF ' (u_j) \frac{y_i}{y_j} .
\end{align*}
Since for $\omega _{jm} \in \Omega_{na}$,  $ \omega _{jm} =
\bar{\omega} _{jm}$, we have
\begin{align*}
	\dot{\omega}_{ij} & = \cF'_j [u_j] \frac{y_i }{y_j} (-\gamma y_j
	\frac{\partial E}{\partial y_j} \\
	& + \sum _{\omega _{jm} \in \Omega_a} \omega _{jm} \
	\dot{\omega}_{jm}+ \sum _{\omega _{jm} \in \Omega_{na}}
	\bar{\omega} _{jm} \ \dot{\omega}_{jm}).
\end{align*}
which is Equation (\ref{Alg}). That is, Equation (\ref{Adap}) is
the unique solution to Equation (\ref{Alg}).

\hfill \bull

Equation (\ref{Alg}) provides us a recursive way to learn gains of
upstream branches from gains and their derivatives of downstream
branches. Therefore, Equation (\ref{Alg}) can be implemented
locally, that is, information needed to learn ${\omega}_{ij}$ is
local to nodes $i$ and $j$ and the branches connected to nodes $i$
and $j$.

Obviously, Equation (\ref{Alg}) can be implemented on-line. So,
the learning can be done on-line as the system runs.

Note that although for $\omega _{jm} \in \Omega_{na}$,  $ \omega _{jm} =	\bar{\omega} _{jm}$ is a constant, we still need to calculate its derivative $\dot{\omega}_{jm}$. However, $\dot{\omega}_{jm}$ is not used to update $\bar{\omega} _{jm}$, but to be used to calculate $\dot{\omega}_{ij}$ as shown in Equation (\ref{Alg}).

Let us consider the following two special cases. \\
{\em Case 1:} Note $j$ is an output node. In this case, Equation
(\ref{Alg}) reduces to
\begin{equation} \label{outputN}
	\begin{split}
		& \dot{\omega}_{ij} = -\gamma y_i \cF'_j [u_j] \frac{\partial
			E}{\partial y_j}.
	\end{split}
\end{equation}
{\em Case 2:} Note $j$ is not an output node. In this case,
Equation (\ref{Alg}) reduces to
\begin{equation} \label{nooutputN}
	\begin{split}
		& \dot{\omega}_{ij} = \cF'_j [u_j] \frac{y_i }{y_j} (\sum _{\omega
			_{jm} \in \Omega_a} \omega _{jm} \ \dot{\omega}_{jm}+ \sum
		_{\omega _{jm} \in \Omega_{na}} \bar{\omega} _{jm} \
		\dot{\omega}_{jm}).
	\end{split}
\end{equation}
Note that the adaptation/learning rate $\gamma$ does not appear in
the above equation.

To ensure Equation (\ref{Alg}) has a unique solution, let us write
Equation (\ref{Alg}) in matrix form as follows. Enumerate the
branches/gains in $\Omega$ as
$$
\Omega = \{\omega ^1, \omega ^2, ..., \omega ^L  \},
$$
where $L = |\Omega|$ is the cardinality (number of elements) of
$\Omega$.

Then, we can write Equation (\ref{Alg}) is the matrix form as
\begin{align*}
	& \left[ \begin{array}{l} \dot{\omega}^1 \\
		\dot{\omega}^2\\
		... \\
		\dot{\omega}^L
	\end{array} \right] =
	\left[ \begin{array}{llll} \phi_{11} & \phi_{12} & ... & \phi_{1L} \\
		\phi_{21} & \phi_{22} & ... & \phi_{2L} \\
		... \\
		\phi_{L1} & \phi_{L2} & ... & \phi_{LL}
	\end{array} \right]
	\left[ \begin{array}{l} \dot{\omega}^1 \\
		\dot{\omega}^2\\
		... \\
		\dot{\omega}^L
	\end{array} \right] +
	\left[ \begin{array}{l} \mu_1 \\
		\mu_2 \\
		... \\
		\mu_L
	\end{array} \right] .
\end{align*}
where $\phi_{lm}$ for $\omega^l= \omega _{ij}$ and $\omega^m=
\omega _{i'j'}$ is defined as follows. If $j \not= i'$, then
$\phi_{lm}=0$; otherwise,
$$
\phi_{lm}=\cF'_j [u_j] \frac{y_i }{y_j} \omega _{jj'}
$$

Denote the above matrix equation as
$$
\dot{\omega} = \Phi \dot{\omega} + \mu,
$$
where $\Phi=[\phi_{ij}]$ is an $L \times L$ matrix and $\omega$
and $\mu$ are column vectors of $L$ dimension.

It is clear that Equation (\ref{Alg}) has a unique solution if and
only if the determinant
$$
|I-\Phi| \not= 0,
$$
where $I$ is the identity matrix of $L \times L$. In the rest of
the paper, we assume that $|I-\Phi| \not= 0$.

Note that we do not need to calculate $\Phi$, $I-\Phi$, or $(I-\Phi)^{-1}$, as Equation (\ref{Alg}) can be implement efficiently using, say, MATLAB/Simulink, as shown in Sections VI and VII. $\Phi$ is introduced only to investigate the uniqueness of solution to Equation (\ref{Alg}).

\section{Fr\'{e}chet Derivatives}

From Equation (\ref{Alg}), it is clear that the key to the
gradient-decent-based on-line learning is the calculation of
Fr\'{e}chet derivatives $\cF'_n [u_n], n \in \cN$. In this
section, we investigate how to calculate Fr\'{e}chet derivatives.

Consider a node $n \in \cN$. If $n$ is not a super node, then $y_n
= u_n$. Hence $\cF'_n [u_n]=1$. If $n$ is a super node, then $y_n
= \cF_n [u_n]$ describes a linear or nonlinear
single-input-single-output system.

We first consider linear systems. A linear system is given either
by a state-space representation or a transfer function. If a
system is given by a state-space representation
\begin{align*}
	& \dot{x} = Ax+Bu \\
	& y=Cx+Du,
\end{align*}
where $x$, $u$, and $y$ are the state variable, input, and output,
respectively, and $A,B,C,D$ are matrices of appropriate
dimensions, then we can convert it into a transfer function
$$
G(s) = C (sI-A)^{-1} B+D.
$$

Let us denote the transfer function of node $n$ by $G_n(s)$ and
its corresponding (unit) impulse response by $g_n(t)$. Then the
dynamics of node $n$ is described by
$$
y_n(t) = \cF _n [ u_n(t)] = g_n(t) * u_n(t),
$$
where $*$ denotes the convolution.

Then, the Fr\'{e}chet derivative of $\cF _n$ can be calculated as
follow.
\begin{equation} \nonumber
	\begin{split}
		& \lim _{||\varepsilon|| \rightarrow 0} \frac{||\cF _n[u+\varepsilon] - \cF _n[u] - \cF' _n[u] \varepsilon||} {||\varepsilon||} = 0 \\
		\Leftrightarrow
		& \lim _{||\varepsilon|| \rightarrow 0} \frac{||g_n(t) *(u_n(t) +\varepsilon) - g_n(t) * u_n(t) - \cF' _n[u] \varepsilon||} {||\varepsilon||} = 0 \\
		\Leftrightarrow
		& \lim _{||\varepsilon|| \rightarrow 0} \frac{||g_n(t) 		*\varepsilon - \cF' _n[u] \varepsilon||} {||\varepsilon||} = 0 .
	\end{split}
\end{equation}

Let us take $\varepsilon$ as a constant (also denoted by
$\varepsilon$). Then
$$
g_n(t) * \varepsilon = p_n(t) \varepsilon .
$$
where $p_n(t)$ is the (unit) step response of the system, which is
the integral of the impulse response.
$$
p_n (t) = \int _0^t g_n(\tau) d\tau .
$$

We then have,
\begin{equation} \nonumber
	\begin{split}
		& \lim _{||\varepsilon|| \rightarrow 0} \frac{||\cF 		_n[u+\varepsilon] - \cF _n[u] - \cF' _n[u] \varepsilon||} {||\varepsilon||} = 0 \\
		\Leftrightarrow
		& \lim _{||\varepsilon|| \rightarrow 0} \frac{||p_n(t) 		\varepsilon - \cF' _n[u] \varepsilon||} {||\varepsilon||} = 0 .
	\end{split}
\end{equation}
Hence
$$
\cF' _n[u] = p_n(t).
$$

Let $t \rightarrow \infty$, then $\cF' _n[u]$ can be approximated as
\begin{equation} \nonumber
	\begin{split}
		\cF' _n[u] = p_n(t) \approx \lim _{t \rightarrow \infty} p_n(t) = \lim _{s \rightarrow 0} s P_n(s) = \lim _{s \rightarrow 0} G_n(s),
	\end{split}
\end{equation}
where $P_n(s)$ is the Laplace transforms of $p_n(t)$. Therefore, the Fr\'{e}chet derivative can be approximated by the corresponding transfer function $G_n(s)$ when $s \rightarrow 0$.

We now consider nonlinear systems. We assume that a nonlinear
system is modeled by nonlinear state and output equations
\begin{align*}
	& \dot{x} = f(x,u) \\
	& y=h(x,u),
\end{align*}
where $f$ and $h$ are nonlinear functions. We linearize the
nonlinear system along a nominal trajectory $u^\circ (t), x^\circ
(t), y^\circ (t)$. We assume that the nominal trajectory is the
solution to the nonlinear state and output equations with initial
conditions $x^\circ (0)$ and input $u^\circ (t)$, that is,
\begin{align*}
	& \dot{x} ^\circ (t) = f(x^\circ (t), u^\circ (t)) \\
	& y^\circ (t) = h(x^\circ (t), u^\circ (t)).
\end{align*}

Define the deviation from the nominal trajectory as
\begin{align*}
	& \Delta x (t) = x (t) - x ^\circ (t)  \\
	& \Delta u (t) = u (t) - u ^\circ (t) ).
\end{align*}
Then the linearized system is given by
\begin{align*}
	& \Delta \dot{x} = A^\circ \Delta x + B^\circ \Delta u \\
	& \Delta y = C^\circ \Delta x + D^\circ \Delta u,
\end{align*}
where $A^\circ , B^\circ , C^\circ , D^\circ$ are the following
Jacobian matrices.
\begin{align*}
	& A^\circ = {\frac{df}{dx}}|_{x=x ^\circ (t), y=y ^\circ (t)}  &
	B^\circ = \frac{df}{du}|_{x=x ^\circ (t), y=y ^\circ (t)} \\
	& C^\circ = \frac{df}{dx}|_{x=x ^\circ (t), y=y ^\circ (t)} &
	D^\circ = \frac{df}{du}|_{x=x ^\circ (t), y=y ^\circ (t)} .
\end{align*}

We find the transfer function of the linearized systems as
$$
G^\circ(s) = C^\circ (sI-A^\circ)^{-1} B^\circ + D^\circ.
$$
The method used above to find Fr\'{e}chet derivative for linear
systems can then be used to find Fr\'{e}chet derivative for
nonlinear systems.

\section{Adaptive PID Control}

In this section, we apply the proposed on-line learning algorithm to model
reference adaptive PID control. The GSFG of the system is shown in
Figure 3, where SN 1 represents the integral part, SN 2 represents
the proportional part, SN3 represents the derivative part, and SN
4 represents the plant to be controlled. The PID gains are given
by $K_I = \omega _{14} , K_P = \omega _{24} , K_D = \omega _{34}$.

\begin{figure}[htb] \label{PID_GSFG}
	\centering
	\includegraphics[keepaspectratio=true,angle=0,width=0.5\textwidth]{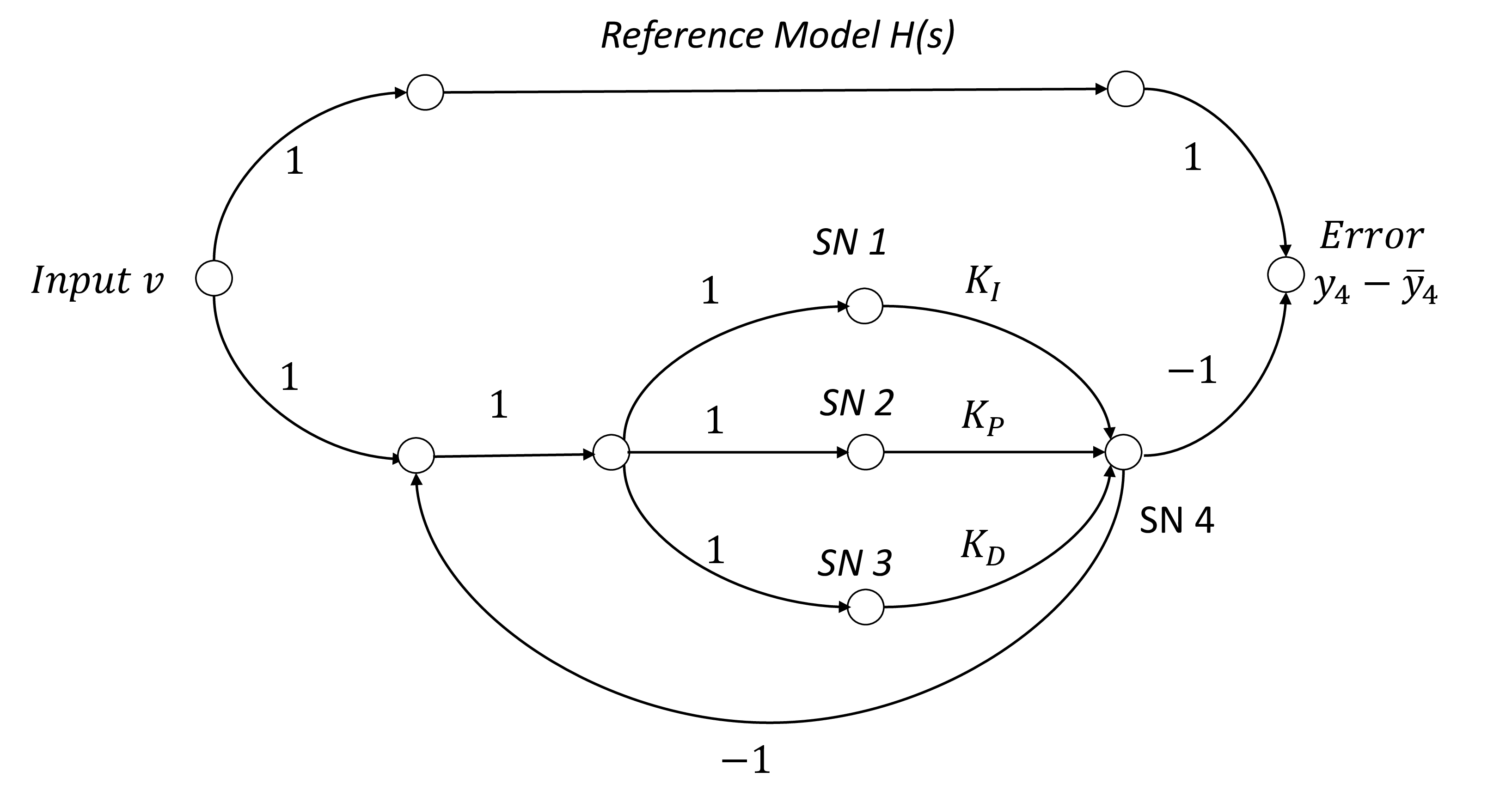}
	\caption{The GSFG of model reference adaptive PID control.}
\end{figure}

The objective is to adapt the gains $K_I, K_P, K_D$ so that the
output of the controlled system follows the output of the
reference model. We assume that the reference model is described
by a transfer function $H(s)$, that is,
$$
\tilde{y}_4 (t)= h(t)*v(t),
$$
where $h(t)$ is the impulse response of the reference model, or
the inverse Laplace transform of $H(s)$, and $v(t)$ is the input
signal being applied to both the reference model and the
controlled system. Hence,
$$
E=\frac{1}{2}(y_4 - \tilde{y}_4)^2.
$$

Since SN 4 is an output node, the on-line learning algorithm is
given by
$$
\dot{\omega}_{i4} = -\gamma y_i \cF'_4 [u_4] \frac{\partial
	E}{\partial y_4} = -\gamma y_i \cF'_4 [u_4] (y_4 - \tilde{y}_4).
$$
In other words,
\begin{equation} \label{PID}
	\begin{split}
		& \dot{K}_I = -\gamma y_1 \cF'_4 [u_4] (y_4 - \tilde{y}_4) \\
		& \dot{K}_P = -\gamma y_2 \cF'_4 [u_4] (y_4 - \tilde{y}_4) \\
		& \dot{K}_D = -\gamma y_3 \cF'_4 [u_4] (y_4 - \tilde{y}_4) .
	\end{split}
\end{equation}

Equation (\ref{PID}) will be used to adapt the PID gains.
\begin{figure*}[htbp] \label{PID_GSFG}
	\centering
	\includegraphics[keepaspectratio=true,angle=0,width=1\textwidth]{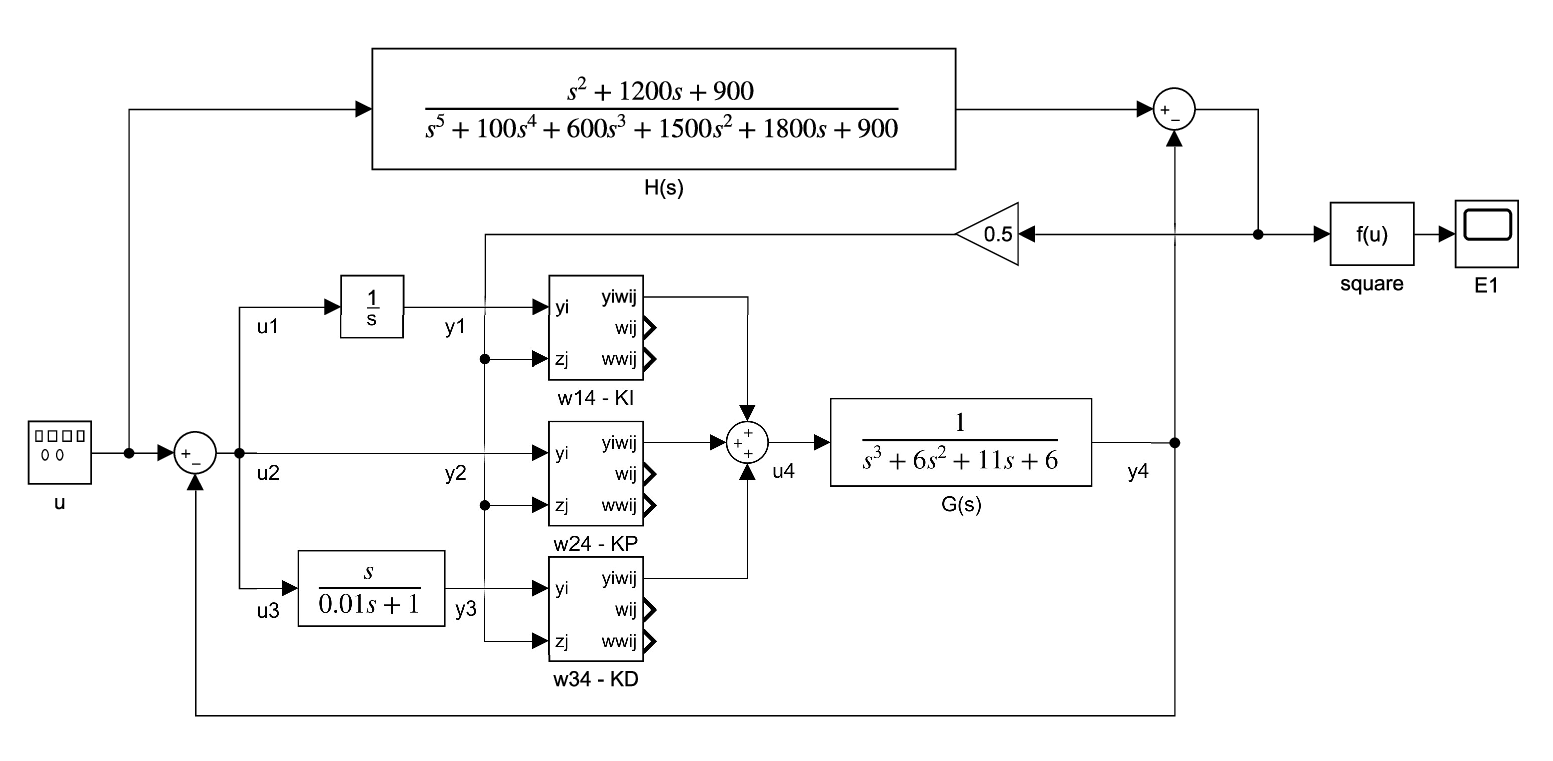}
	\caption{The Simulink implementation of model reference adaptive
		PID control.}
\end{figure*}

\section{Simulation Results}

We implemented the above model reference adaptive PID control in
Simulink as shown in Figure 4. In the figure, the system at the
top is the reference model.

In all simulations, we use the same reference model with the
following transfer function.
$$
H(s) = \frac{s^2+ 1200 s +900}{s^5 + 100 s^4 + 600 s^3 + 1500 s^2
	+ 1800 s + 900} .
$$
The reference model $H(s)$ is selected to have good transient
response, whose step response is shown in Figure 5. The poles of
$H(s)$ are
\lfteqn
p_1= -93.7698, & \\
p_2 = -1.7941+j0.7362, & p_3 = -1.7941-j0.7362, \\
p_4 = -1.3210+j0.8984, & p_5 = -1.3210-j0.8984 .
\ndeqn

The system at the bottom is the controlled system controlled by
the adaptive PID controller. The plant to be controlled is at the
right and the adaptive PID controller at the left. Three adaptive
blocks implement the Equation (\ref{PID}).

\begin{figure}[h] \label{RM}
	\centering
	\includegraphics[keepaspectratio=true,angle=0,width=0.48\textwidth]{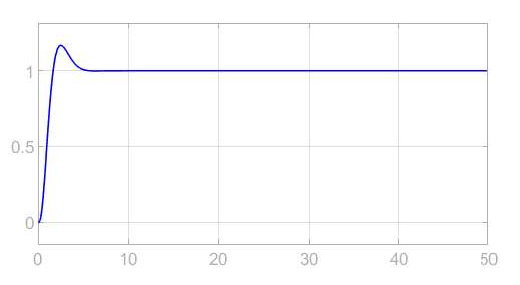}
	\caption{Step response of the reference model $H(s)$.}
\end{figure}

The same input signals are applied to both the reference model and
the controlled system. We simulate the system using different
input signals (square, sawteeth, and sinusoidal) and find that
they all work well.

Various simulations are performed for different systems. In the
simulations, we do not attempt to find the optimal $\gamma$,
rather, we test several values for $\gamma$ and then pick one that
works well. The simulations results of some typical systems are
presented below. 

In the simulations, the initial values of $K_P, K_I, K_D$ are selected rather arbitrarily as
$$
K_P=12, \ \  K_I=8, \ \ K_D=4.
$$ 
We selected other initial values as well. The simulations show that the selection of the initial values are not important.

~\\

\noindent {\bf Stable Plant}

we start our simulation with a linear stable plant having the
following transfer function
$$
G_1(s) = \frac{1}{s^3 + 6s^2 + 11 s +6} .
$$
It is easy to check that the above plan is stable. The simulation
results show that the adaptive PID controller works well and the
error approaches 0 and $K_P, K_I, K_D$ converges as shown in
Figure 5. 

~\\

\begin{figure}[h] \label{RM}
	\centering
	\includegraphics[keepaspectratio=true,angle=0,width=0.48\textwidth]{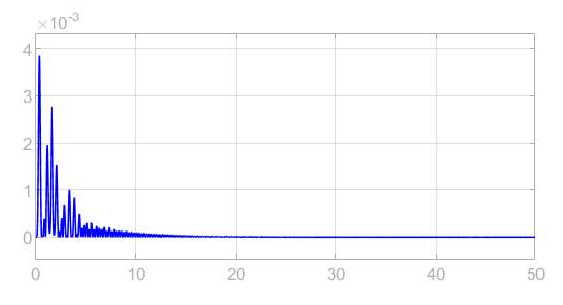}
	\includegraphics[keepaspectratio=true,angle=0,width=0.48\textwidth]{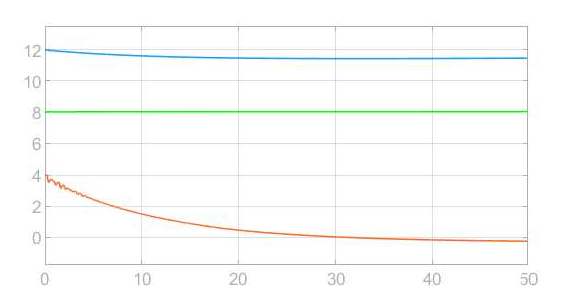}
	\caption{Simulation results for a stable plant. Top figure shows the error $E$ and the bottom shows $K_P$ (blue), $K_I$ (green), and $K_D$ (coral).}
\end{figure}

\noindent {\bf Unstable Plant}

We then simulate a plant with the following transfer function
$$
G_2(s) = \frac{1}{s^3 + 6s^2 + 11 s -6} .
$$
It is easy to check that the above plan is unstable. Even thought,
the simulation results show that the adaptive PID controller works
well and the error approaches 0 and $K_P, K_I, K_D$ converges as
shown in Figure 6. 

\begin{figure} \label{RM}
	\centering
	\includegraphics[keepaspectratio=true,angle=0,width=0.48\textwidth]{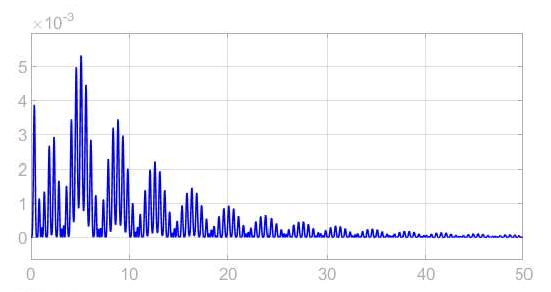}
	\includegraphics[keepaspectratio=true,angle=0,width=0.48\textwidth]{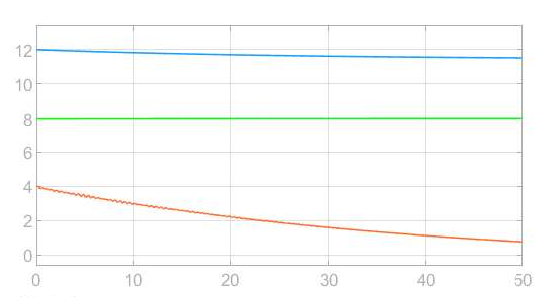}
	\caption{Simulation results for unstable plant. Top figure shows
		the error $E$ and the bottom shows $K_P$ (blue), $K_I$ (green),
		and $K_D$ (coral).}
\end{figure}

From Figure 6, we see that if the plant is unstable, then the closed-loop system is unstable initially. This is because we select the initial values of $K_P, K_I, K_D$ rather arbitrarily without the need of considering stability. Hence, the error grows at the first 5 seconds. Then, as the PID controller adapts, the closed-loop system becomes stable around 5 seconds. The PID controller continues to adapt and then error becomes smaller and smaller.

~\\

\noindent {\bf Systems with Time Delay}

The learning algorithm also works for systems with time delay. By
adding a delay of 0.03 second before the plant with transfer
function $G_1(s)$, we obtain the simulation results as shown in
Figures 7. Clearly, the error approaches 0. 

~\\

\begin{figure} \label{RM}
	\centering
	\includegraphics[keepaspectratio=true,angle=0,width=0.5\textwidth]{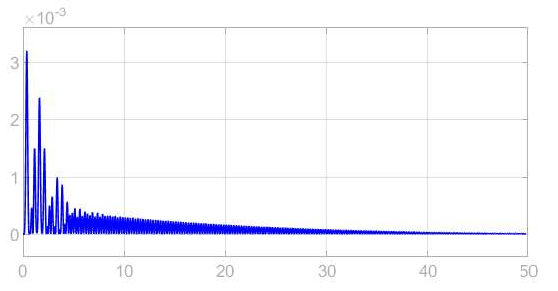}
	\includegraphics[keepaspectratio=true,angle=0,width=0.5\textwidth]{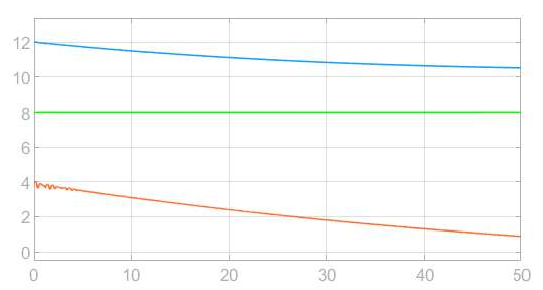}
	\caption{Simulation results for plant with time delay. Top figure
		shows the error $E$ and the bottom shows $K_P$ (blue), $K_I$
		(green), and $K_D$ (coral).}
\end{figure}

\noindent {\bf Nonlinear Systems}

If the plant is nonlinear, the learning algorithm also works.
Figures 8 shown simulation results of a nonlinear plant with the
following dynamics.
\lfteqn
\dot{x}_1= - x_1 + 0.5 \sin (x_1) +u_4 \\
\dot{x}_2= - 2 x_2 -x_2^3 +x_1 \\
\dot{x}_3= - 3 x_3 - 0.2 \tan (x_3) +x_2 \\
y_4 = x_3 .
\ndeqn

\begin{figure} \label{RM}
	\centering
	\includegraphics[keepaspectratio=true,angle=0,width=0.5\textwidth]{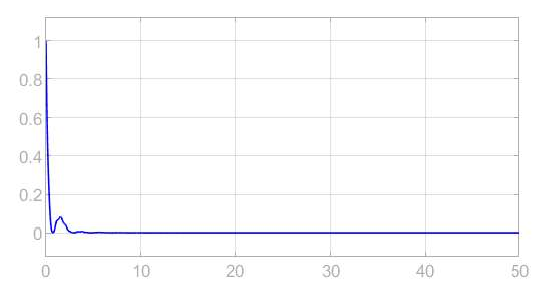}
	\includegraphics[keepaspectratio=true,angle=0,width=0.5\textwidth]{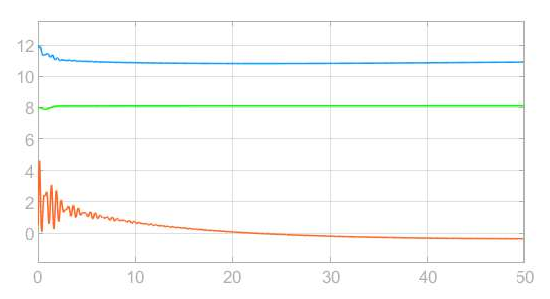}
	\caption{Simulation results for nonlinear plant. Top figure shows
		the error $E$ and the bottom shows $K_P$ (blue), $K_I$ (green),
		and $K_D$ (coral).}
\end{figure}

The above simulations show that our adaptation law works very well for different types of system.

\section{Conclusion}

In this paper, we extend the Brandt-Lin learning algorithm for
neural networks to dynamics systems and apply it to model
reference adaptive control. The main contributions of the paper
are as follows. (1) Generalized conventional signal-flow graphs to
model general dynamic systems with both linear and nonlinear
dynamics. (2) Derive the extended Brandt-Lin algorithm and the
necessary and sufficient condition for its unique solution. (3)
Apply the extended Brandt-Lin algorithm to model reference
adaptive control and derive adaptation law for adaptive PID
controllers. We plan to apply the results to other adaptive
controllers in the future.

\bibliographystyle{ieeetr}

\bibliography{PID}             

\end{document}